# In-situ crosslinked wet spun collagen triple helices with nanoscale-regulated ciprofloxacin release capability

M. Tarik Arafat,[1,2,§] Giuseppe Tronci,[1,2,*] David J. Wood,[2] Stephen J. Russell[1]

[1]Clothworkers' Centre for Textile Materials Innovation for Healthcare, School of Design, University of Leeds, UK

[2]Biomaterials and Tissue Engineering Research Group, School of Dentistry, St. James University Hospital, University of Leeds, UK

[§] Present address: Department of Biomedical Engineering, Bangladesh University of Engineering and Technology, Dhaka, Bangladesh

[*] Email correspondence: g.tronci@leeds.ac.uk (G.T.)

## Highlights

- Cip-encapsulated, in-situ crosslinked wet spun collagen fibres were accomplished
- In-situ Ph-crosslinked fibres displayed the highest tensile modulus
- Nanoscale aromatic interactions proved key to control Cip release
- Fibre morphology was not affected by drug encapsulation and in-situ crosslinking

## Abstract

The design of antibacterial-releasing coatings or wrapping materials with controlled drug release capability is a promising strategy to minimise risks of infection and medical device failure *in vivo*. Collagen fibres have been employed as medical device building block, although they still fail to display controlled release capability, competitive wet-state mechanical properties, and retained triple helix organisation. We investigated this challenge by pursuing a multiscale design approach integrating drug encapsulation, in-situ covalent crosslinking and fibre spinning. By selecting ciprofloxacin (Cip) as a typical antibacterial drug, wet spinning was selected as a triple helix-friendly route towards Cip-encapsulated collagen fibres; whilst in-situ crosslinking of fibre-forming triple helices with 1,3-phenylenediacetic acid (Ph) was hypothesised to yield Ph-Cip π-π stacking aromatic interactions and enable controlled drug release. Higher tensile modulus and strength were measured in Ph-crosslinked fibres compared to state-of-the-art carbodiimide-crosslinked controls. Cip-encapsulated Ph-crosslinked fibres revealed decreased elongation at break and significantly-enhanced drug retention *in vitro* with respect to Cip-free variants and carbodiimide-crosslinked controls, respectively. This multiscale manufacturing strategy provides new insight aiming at wet spun collagen triple helices with nanoscale-regulated tensile properties and drug release capability.

**Graphical abstract**

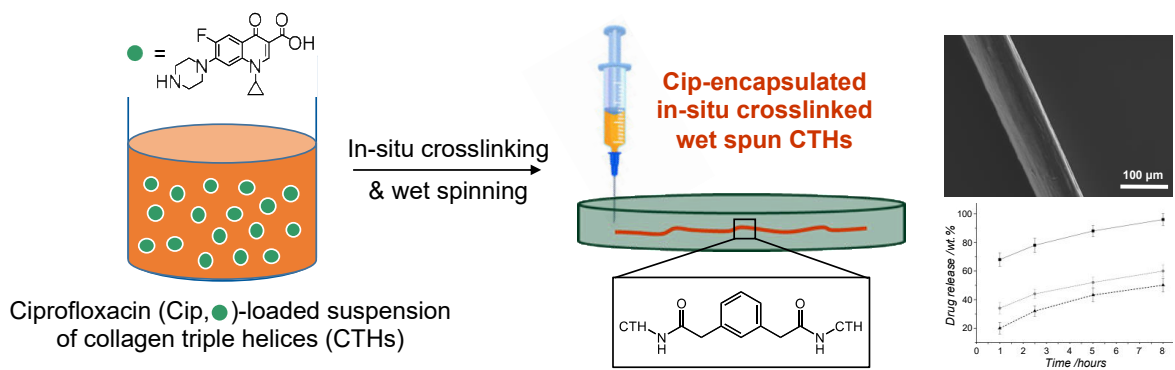

## 1. Introduction

Bacterial infection is one of the most common causes of medical implant failure. Pathogenic micro-organisms are found in approximately 90% implants [1], potentially leading to bone infection, such as osteomyelitis. Here, the use of fluoroquinolones is an established systemic therapeutic approach. Ciprofloxacin (Cip) has been the most widely used fluoroquinolone to treat bacterial bone infection, due to its low (0.25–2 µg/mL) minimal inhibitory concentration (MIC) for most osteomyelitis-related pathogens [2]. The design of a medical device coating or bone wrap capable of delivering Cip in a controlled and localised manner is therefore an appealing strategy to minimise risks of osteomyelitis.

Collagen-based fibres have been widely manufactured for biomedical applications due to their inherent biomimetic features at multiple length scales [3]. In light of collagen's hierarchical organisation, bespoke fibre spinning approaches have successfully been developed to realise mechanically-competitive wet spun collagen fibres [4], highly ordered structures [5], and nonwoven collagen-based assemblies [6]. These building blocks could be applied as delivery system of antibacterial drugs or used to wrap existing medical devices, delivering a cost-effective infection control strategy. In light of the high compatibility with water, however, collagen-based fibres often display limited structural stability [7] and drug delivery capability [8], whereby antibacterial drugs, e.g. Cip, are released fast, typically within minutes [9]. To achieve durable antibacterial release and minimise risks of infection-induced failure of medical devices, multiscale mechanisms of collagen fibre spinning and drug-fibre complexation are key.

This study investigated the design of Cip-encapsulated fibres via a multiscale integrated process of collagen triple helix (CTH) wet spinning and in-situ crosslinking. We hypothesised that nanoscale aromatic interactions could be introduced between fibre-forming covalently-crosslinked collagen triple helices and Cip, aiming at enhanced fibre spinnability, mechanical compliance and drug delivery capability. In-situ crosslinking was pursued during wet spinning of Cip-encapsulated collagen suspensions via either CTH functionalisation with 1,3-phenylenediacetic acid (Ph) or control carbodiimide-mediated condensation reaction. Resulting wet spun fibres were assessed with respect to lysine functionalisation, surface morphology, tensile properties, drug release capability and hydrolytic degradability.

## 2. Materials and methods

### 2.1 Materials

1-ethyl-3-(3-dimethylaminopropyl) carbodiimide hydrochloride (EDC), N-hydroxysuccinimide (NHS), Ph, and β-mercaptoethanol (βME) were purchased from Alfa Aesar. Cip was purchased from Cambridge Bioscience. 2,4,6-trinitrobenzenesulfonic acid (TNBS), acetic acid and Dulbecco's Phosphate Buffered Solution (PBS) were purchased from Sigma Aldrich.

### 2.2 Drug encapsulation, in-situ crosslinking and wet spinning

In-house extracted rat tail collagen suspensions were prepared (1.2 % wt./vol.) in 17.4 mM acetic acid [4]. Three in-situ crosslinking strategies were pursued: (i) Ph- and EDC-mediated network formation, (ii) Ph-mediated network formation, and (iii) state-of-the-art carbodiimide-mediated condensation reaction control. In (i), Ph (4 mg) was dissolved in the collagen suspension (3 ml) and an equimolar amount (0.01 M) of EDC and NHS was added prior to wet spinning [4]. In (ii), Ph, EDC and NHS were added as previously reported, whilst an equimolar amount of βME (with respect to EDC) was added to quench EDC and stirred for 30 min prior to wet spinning. In (iii), an equimolar content (0.01 M) of EDC and NHS was dissolved in the collagen suspension prior to wet spinning. Cip-encapsulated fibres were obtained according to (i-iii), whereby 15 μg·ml$^{-1}$ Cip was added to the collagen suspension prior to addition of the crosslinking agents. Crosslinking occurrence was assessed via TNBS assay (n=3) [10].

**2.3 Tensile tests and scanning electron microscopy of wet spun collagen fibres**

Tensile tests were carried out (0.03 mm·s$^{-1}$, 18 ℃, 38 % r.h.) on individual collagen fibres (n=10) using a Zwick Roell Z010 apparatus equipped with a 10 N load cell. Tensile measurements were reported as mean ± standard deviation. Fibre surface morphology was inspected using a Hitachi SU8230 FESEM, with a beam intensity of 10 kV after gold sputtering using a JFC-1200 fine sputter coater.

**2.4 Drug release and degradation tests**

In-situ crosslinked wet spun fibres (n=3) of known weight were individually stored in centrifuge tubes containing 5 ml PBS at 37 ℃. Cip release was quantified at selected time points by measuring the supernatant absorbance at 330 nm (U-3010, Hitachi High-Technologies Corporation, Japan). Following 21-day incubation in PBS, fibres were rinsed with distilled water, dried and weighed, to quantify any sample mass loss [4].

## 3. Results and discussion

A multiscale integrated process of Cip encapsulation, CTH in-situ crosslinking and wet spinning was developed aiming at mechanically-competent fibres with antibacterial functionality for medical devices (Figure 1).

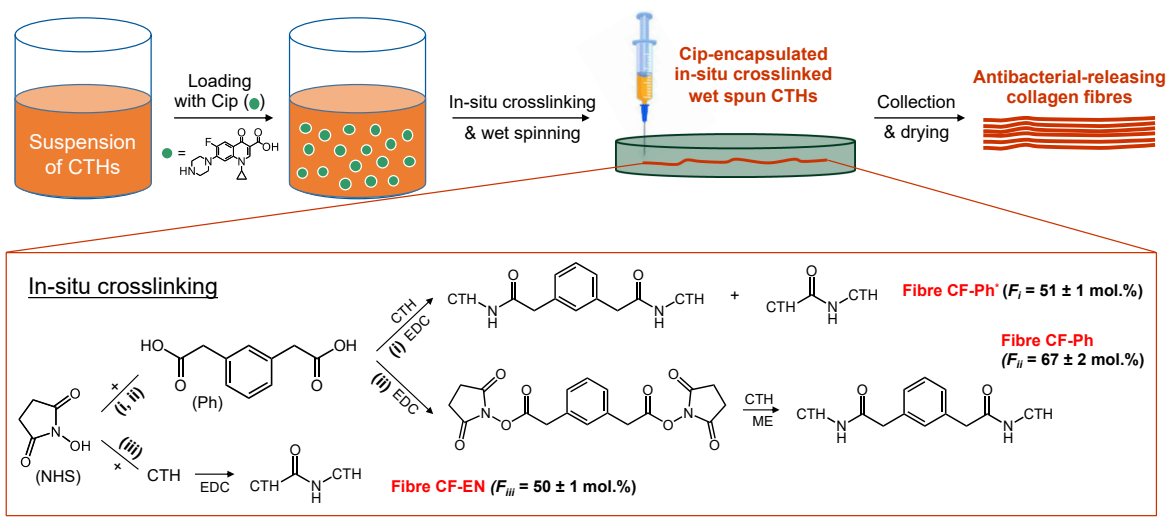

**Figure 1.** Multiscale integrated process yielding mechanically-competent fibres as antibacterial-releasing wrapping material for medical devices. In-situ crosslinking of wet spinning CTHs was pursued via reaction with NHS-activated Ph in the absence (i) or presence (ii) of ME, and via control EDC-induced condensation reaction (iii). Ph was selected to mediate $\pi$-$\pi$ aromatic interactions with Cip and to control fibre's drug release capability.

In-situ crosslinking was pursued during wet spinning to enhance the spinnability of CTH suspension, enable homogeneous crosslink density in fibre-forming CTHs, and control fibre's release capability via Ph-Cip $\pi$-$\pi$ aromatic interactions. NHS-activated Ph was reacted with CTHs in the absence (i) or presence (ii) of βME, whilst carbodiimide-induced crosslinking reaction (iii) was carried out as state-of-the-art reaction control. Therefore, covalent networks of CTHs crosslinked with (i) both EDC-induced amide bonds and Ph-based aromatic junctions, (ii) Ph-based aromatic junctions only, and (iii) EDC-induced amide bonds only (iii), were expected in resulting fibres.

Wet spun fibres were successfully realised with all three in-situ crosslinking strategies, whereby an averaged degree of collagen functionalisation ($F$) was measured via TNBS assay in the range of 50-67 mol.% (Figure 1). CF-Ph$^*$ obtained via (i) displayed comparable $F$ with respect to EDC-crosslinked collagen fibre controls (CF-EN) deriving from (iii); whilst the highest value of $F$ was recorded in fibres CF-Ph crosslinked via (ii) with Ph-based aromatic junctions only. Given that comparable molar crosslinker ratios were used in (i-iii), above-mentioned $F$ trends reflect selected crosslinking mechanisms. Reaction of collagen with NHS-activated Ph leads to lysine functionalisation with an aromatic residue and increased probability of crosslinking distant CTHs [11]; whilst zero-length crosslinks are generated via EDC-induced crosslinking reaction control [12], so that steric effects play a major role on the crosslinking yield. When Ph-mediated in-situ crosslinking of CTHs is carried out concomitantly to EDC-induced condensation reaction (in the absence of βME), the competition between theSE two reactions is likely to explain the reduced degree of functionalisation in fibres CF-Ph$^*$ with respect to fibres CF-Ph.

Other than the molecular scale, resulting fibres displayed uniform surface morphology under SEM (Figure 2 A). A diameter of ~90 µm was measured, whereby minimal surface and diameter variations were observed with respect to the case of non-crosslinked, Cip-free fibre obtained via wet spinning of the same collagen suspension (Figure 2 B). This observation provides evidence that neither the introduction of a covalent network at the molecular scale or the encapsulation with Cip impacted on wet spun fibre microscale.

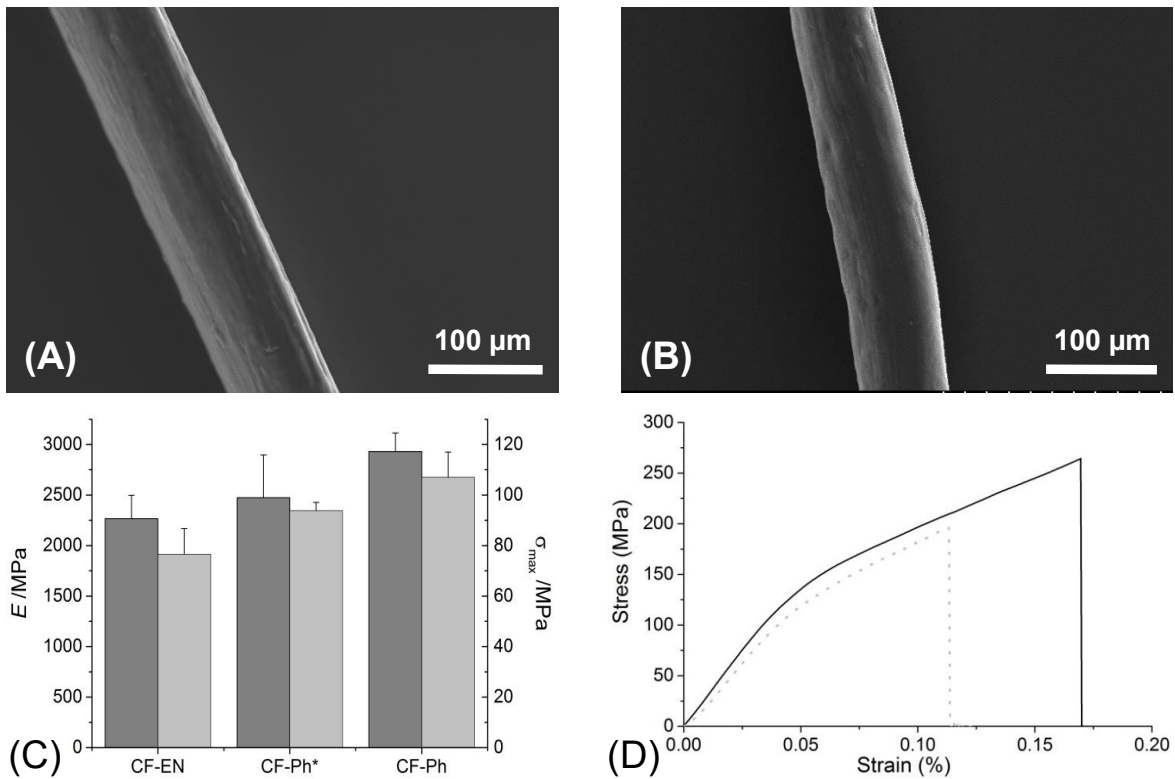

**Figure 2.** (A-B): SEM images of fibre Cip-CF-Ph (A) and non-crosslinked Cip-free control (B). (C): Tensile modulus ($E$, grey) and maximal tensile stress ($\sigma_{max}$, light grey) of Cip-encapsulated in-situ crosslinked wet spun collagen fibres. (D): Stress-strain curves of fibres CF-Ph with (⋯) and without (—) Cip encapsulation.

At the macroscale, the tensile modulus ($E$) and maximal tensile stress ($\sigma_{max}$) of the Cip-encapsulated in-situ crosslinked fibres were measured in the range of 2100-2900 MPa and 75-100 MPa, respectively (Figure 2 C). Samples Cip-CF-Ph revealed the highest averaged values ($E$: 2824 MPa; $\sigma_{max}$: 98 MPa), which proved to be statistically different to those of the control group Cip-CF-EN. The higher values of tensile properties measured in fibres Cip-CF-Ph compared to controls Cip-CF-EN and, to a lesser extent, to fibres Cip-CF-Ph$^*$, reflects the above-mentioned considerations in the network configuration introduced at the molecular scale of the wet spun fibres (Figure 1). Intermolecular crosslinks are selectively formed in the former samples following nucleophilic addition of collagen amine terminations with activated Ph carboxylic groups, in contrast to the latter groups [4,11,12]. The superior tensile mechanical properties of the Ph-crosslinked fibres are also supported by the increased degree of functionalisation in samples CF-Ph compared to controls CF-EN (Figure 1), indirectly suggesting the formation of covalent networks with increased crosslink density and $\pi$-$\pi$ aromatic interactions between Ph-functionalised CTHs [11]. When comparing Cip-free and Cip-encapsulated fibres CF-Ph, a significantly decreased strain at break ($\varepsilon_b$) and insignificantly lower tensile stress were recorded in the latter compared to the former group (Figure 2 D). These results could be due to the Cip-induced plasticising effect of resulting fibres, due to the establishment of Cip-Ph rather than Ph-Ph $\pi$-$\pi$ aromatic stacking interactions.

Figure 3 (A) reports the temporal release profile of Cip following fibre incubation in PBS. Significant differences in fibre release capability were observed depending on the in-situ crosslinking strategy adopted during fibre wet spinning. In situ CTH crosslinking via simultaneous Ph-induced functionalisation and control EDC-mediated condensation reactions, on the one hand, and selective Ph-induced functionalisation reaction, on the other hand, successfully led to collagen fibres with increased temporal retention of Cip, indirectly supporting the establishment of $\pi$-$\pi$ aromatic interactions between Ph-crosslinked CTHs and Cip. In the first hour, both Ph-reacted samples showed reduced release of Cip by more than 50 % compared to the Ph-free control group. Within 8 hours, control fibres Cip-CF-EN displayed nearly complete release of Cip, whereas fibres Cip-CF-Ph[*] and Cip-CF-Ph still proved to retain ~40 % and ~50 % of the drug, respectively. Overall, the release rate was found to be decreased following 2.5-hour incubation of all groups, likely due to the initial water-induced swelling of the dry collagen fibres and diffusion of Cip molecules coating the fibre surface.

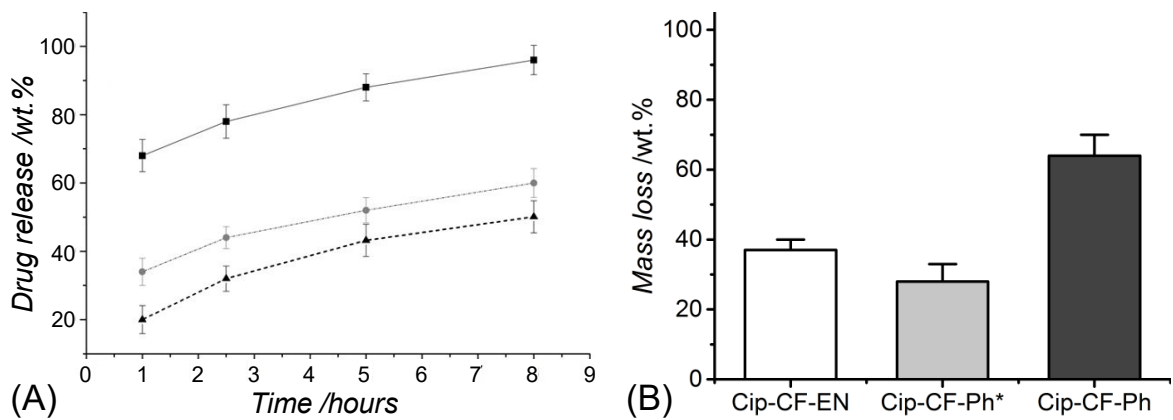

**Figure 3.** (A): Cip release profiles recorded following incubation of Cip-encapsulated fibres *in vitro* (PBS, 37 ºC). (—■—): Cip-CF-EN; (--●--): Cip-CF-Ph*; (--▲--): Cip-CF-Ph. (B): Mass loss measured following 21-day incubation of Cip-encapsulated fibres *in vitro* (PBS, 37 ºC).

Other than the release capability, the hydrolytic degradability of collagen fibres was also investigated by gravimetric analysis (Figure 3 B). Samples CF-Ph described the highest mass loss, followed by control samples CF-EN and fibres CF-Ph$^*$, a result which appeared to be in disagreement with the increased degree of collagen functionalisation (Figure 1) and tensile properties (Figure 2 C) of the former compared to the latter samples. The most likely explanation for this finding is that the addition of βME soon after Ph activation (Figure 1, ii) shifted the functionalisation reaction towards grafting rather than crosslinking, so that an increased content of Ph-grafted CTHs was generated at the molecular scale of fibres CF-Ph.

## 4. Conclusions

A multiscale integrated process of ciprofloxacin encapsulation, in-situ crosslinking, and wet spinning was successfully developed to realise collagen fibres as medical devices with antibacterial release functionality. Fibre's Cip release capability proved to be regulated at the nanoscale via the establishment of $\pi$-$\pi$ aromatic interactions between the encapsulated drug and aromatised and crosslinked CTHs. In-situ crosslinked wet spun collagen fibres revealed homogeneous surface morphology with a diameter of ~90 µm, similarly to the case of non-crosslinked wet spun controls. The highest tensile modulus and strength were measured in wet spun fibres selectively crosslinked with Ph, in line with the increased degree of collagen functionalisation.

## Acknowledgments

The Authors gratefully acknowledge the EPSRC Centre for Innovative Manufacture in Medical Devices (MeDe Innovation) and the Clothworkers' Centre for Textile Materials Innovation for Healthcare for funding this work.